\documentclass[9pt,twocolumn,twoside]{pnas-new}

\usepackage[version=3]{mhchem}

\templatetype{pnasresearcharticle} 
\setboolean{displaywatermark}{false}

\begin{document}

\title{Unifying Chemical and Electrochemical Thermodynamics of Electrodes}

\author[a]{Archie Mingze Yao}
\author[b]{Amal Sebastian}
\author[a,b,1]{Venkatasubramanian Viswanathan}

\affil[a]{Department of Mechanical Engineering, University of Michigan, Ann Arbor, MI 48109}
\affil[b]{Department of Aerospace Engineering, University of Michigan, Ann Arbor, MI 48109}

\leadauthor{Yao}



\significancestatement{Thermodynamics of electrodes is the prerequisite to design electrochemical energy storage systems. Gibbs free energy of the electrode materials determines the maximum amount of electrical work available from the system. However, rigorous thermodynamic analysis of electrode materials is rare in battery modeling. Meanwhile, calculation of phase diagram (CALPHAD) style thermodynamic modeling rarely uses electrochemical data, which contain rich information about chemical potential and therefore phase equilibrium of materials. Leveraging differentiable programming, this work models thermodynamics of electrodes with electrochemistry and phonon density-of-states data together with traditional thermochemistry and phase equilibrium data, resulting in second-law-consistent open-circuit voltage for battery modeling as well as broadened choice of data for thermodynamic modeling, facilitating cross-talks between these two seemingly distinct but closely related fields.}

 
\authorcontributions{V.V. and A.M.Y conceived the idea for the project, A.M.Y. derived expressions, implemented algorithms and performed research, A.S. proposed the use of orthonormal polynomials, and all authors jointly analyzed the results and wrote the paper}
\authordeclaration{Authors declare no competing interest.}
\correspondingauthor{\textsuperscript{1}To whom correspondence should be addressed. E-mail: venkvis@umich.edu}


\begin{abstract}
Batteries are critical for electrified transportation and aviation, yet thermodynamic understanding of electrode materials remains lacking, as indicated by the often-seen violation of the second law of thermodynamics of open-circuit voltage (OCV) models. On the other hand, thermodynamic modeling rarely utilizes electrochemical data such as OCV, entropic heat (dOCV/dT), which contains rich thermodynamic information. This work introduces a framework of thermodynamic modeling of materials for electrochemical energy storage, using differentiable programming and gradient-based optimization of thermodynamic parameters. Using a modified Debye model that accounts for the phonon density of states, the thermodynamics of pure substances is modeled from experimental measurements of specific heat ($c_p$) as well as the phonon density of states $g(\omega)$. Thermodynamics of mixing is modeled with measured entropic heat and OCV data. We demonstrate the differentiable thermodynamic modeling framework with forward and inverse problems. In the forward problem, i.e. determining phase diagram of graphite anode given thermochemical and electrochemical data, we show that in addition to accurate reproduction of phase diagram of \ce{Li_x C6}, the fitted temperature-dependent OCV of graphite reaches 3.8 mV mean absolute error (MAE) for test set data measured at 10$^\circ$C, compared with 2.9-3.6 mV MAE for training set data measured at 25$^\circ$C - 57$^\circ$C. In the inverse problem, i.e. determining OCV of lithium iron phosphate (LFP) cathode from phase diagram constrained by thermochemical and electrochemical data, we demonstrate accurate reproduction of LFP OCV as well as phase diagram. This framework offers a unified treatment of chemical and electrochemical thermodynamic data for electrode materials.
\end{abstract}

\dates{This manuscript was compiled on \today}
\doi{\url{www.pnas.org/cgi/doi/10.1073/pnas.XXXXXXXXXX}}

\maketitle
\thispagestyle{firststyle}
\ifthenelse{\boolean{shortarticle}}{\ifthenelse{\boolean{singlecolumn}}{\abscontentformatted}{\abscontent}}{}



\dropcap{T}he success of electrified transportation and aviation is highly dependent on electrochemical energy storage systems (e.g. batteries), which supplies the stored energy as electrical work during discharging.\cite{Fredericks2018, Sripad2021_and_chart, Viswanathan2022, pnas_perspective_jcesr} Ultimately, it is the Gibbs free energy of the electrodes that decides the maximum electrical work available from a battery.\cite{Chen2022_textbook} Proper understanding of the thermodynamics of electrodes can also render insights such as deformation mechanism\cite{Zhao2011_Zhigang, Brassart2013_Zhigang}, optimal shape and size of electrode particles that suppress phase change,\cite{Luo2021, Luo2022} which ultimately improves electrode performance.\cite{Luo2021, Luo2022} Thermodynamics of electrode materials in electrochemical systems is typically viewed  as a fitting function. For example, one of the most important thermodynamic properties of electrode materials is the open-circuit voltage (OCV), which is the steady-state voltage when no current flows through a battery and is a fundamental thermodynamic property of the electrodes of lithium-ion batteries.\cite{newman2021electrochemical} OCV models that are not constrained by thermodynamics are commonly found in the battery modeling literature, and as a consequence such OCV models risk violating the second law of thermodynamics.\cite{yao2024open} As the first step of modeling thermodynamics of electrode materials, Karthikeyan et al derived expression of OCV of electrode materials and expanded activity coefficient (i.e. excess Gibbs free energy) with various models.\cite{KARTHIKEYAN20081398} However, the monotony condition of OCV as a function of Li filling fraction as proposed in Yao and Viswanathan\cite{yao2024open} was overlooked in the work and as a consequence, several of the fitted OCV models still violated the second law of thermodynamics. The current state of OCV models implies significant limitations and inconsistencies of our understanding towards the thermodynamics of electrode materials in electrochemical systems, highlighting an urgent need for improvement.  

Thermodynamic modeling provides a rigorous approach to analyzing the thermodynamic properties of materials, resulting in accelerated discovery and design of advanced functional materials.\cite{YiWang2019ICME}. In the classical calculation-of-phase-diagram (CALPHAD)-type thermodynamic modeling, Gibbs free energy expressions of different phases of materials can be inferred from thermochemical data such as measured heat capacity, enthalpy and entropy, first-principle calculated energy data, as well as phase equilibrium data such as phase stability and compositions,\cite{Liu2009} and have been applied to understand thermodynamics of various systems including various alloys\cite{Gasior1996_LiMg, WOLVERTON20022187, Ozturk2005_LongQing_ZiKui_Wolverton}, oxides\cite{Li2019_oxides_inLIB_calphad_review} etc. Some previous work on CALPHAD modeling used electrochemical data such as experimentally measured OCV to verify the accuracy of the modeled Gibbs free energy expressions,\cite{Abe2011, Chang2013, Chang2012} as summarized in Li et al.\cite{Li2019_oxides_inLIB_calphad_review} However, despite having rich thermodynamic information, electrochemical measurements data is seldom used to reconstruct Gibbs free energy in existing literature on thermodynamic modeling. Some recent thermodynamic modeling of electrochemical systems includes Guan et al\cite{P2_PNAS_paper} who analyzed the pressure-dependent electrochemical phase diagram utilizing thermodynamic modeling. Lund et al reported a machine-learning assisted phase diagram generation framework and applied it to discuss the phase diagrams of cathode active materials of Li-ion batteries, in which the open-circuit voltage profile of lithium cobalt oxide is used to optimize the generated phase diagram.\cite{Lund2022} Kim et al reported a electrochemical-thermodynamic measurement of solvation shell energy of electrolyte by measuring OCV of cells with symmetric electrodes and asymmetric electrolytes.\cite{Kim2021_YiCui} However, thermodynamic modeling for electrochemical systems and especially active materials is still a rarity in general. 

\begin{figure}[!ht]
    \centering
    \includegraphics[width=8.7cm]{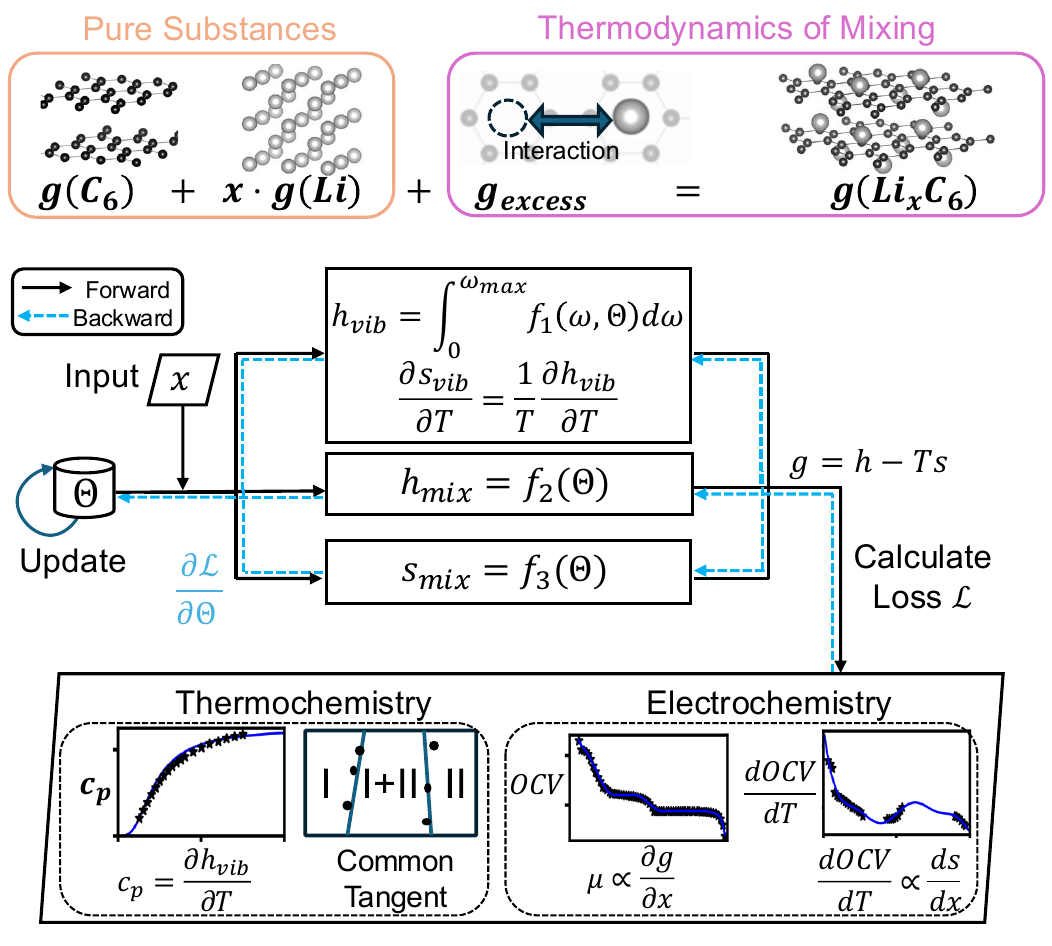}
    \caption{Overview of this work. Thermodynamics of pure substances are modeled by the measured constant pressure specific heat, then the excess mixing Gibbs free energy is modeled by fitting mixing entropy to entropic heat data and fitting mixing enthalpy to OCV data. Finally, the Gibbs free energy of mixed substances can be calculated by combining the pure substance Gibbs free energy and excess Gibbs free energy. Differentiable thermodynamic modeling implemented in Yao and Viswanathan \cite{yao2024open} was applied and extended to determine the parameter values. }
    \label{fig:workflow}
\end{figure}

In this work, we demonstrate a thermodynamic modeling framework of electrode materials of electrochemical energy storage systems as shown in Figure \ref{fig:workflow}, supported by differentiable thermodynamics implemented in differentiable programming frameworks that allows gradient-based  optimization of parameters in the free energy expression. Lying at the core of the thermodynamic modeling is the Gibbs free energy expression of electrode materials, which is decomposed as vibrational enthalpy and entropy contributed by phonon modes for pure substances, and mixing enthalpy and entropy for mixed substances, i.e. lithiated host materials. The expressions of vibrational enthalpy and entropy are derived from statistical thermodynamics and fitted with experimental measured constant pressure specific heat capacity as well as phonon density of states. The mixing entropy of lithiated host material is fitted with the measured entropic heat ($\frac{dOCV}{dT}$). Then, mixing enthalpy can be determined by measured OCV. We demonstrate the differentiable thermodynamic modeling framework with the forward and inverse problems. In the forward problem, i.e. determining phase diagram of graphite anode given thermochemical and electrochemical data, besides the accurately reproduced phase diagram of \ce{Li_x C6}, we extrapolated the fitted temperature-dependent OCV of graphite to low temperature (10 $^\circ$C) that is not in the training data, the graphite OCV model still reaches 3.8 mV mean absolute error (MAE) compared with 2.9-3.6 mV  (MAE) at 25$^\circ$C - 57$^\circ$C. In the inverse problem, we determine OCV of lithium iron phosphate (LFP), one of the most popular cathode materials, by fitting to phase equilibrium data in addition to thermochemical and electrochemical data in fully differentiable manner, resulting in accurate reproduction of LFP phase diagram and a fitted LFP OCV with 0.0276 V MAE, better than the LFP OCV reported in our previous work\cite{yao2024open} while ensuring thermodynamic consistency.


\section*{Thermodynamic Modeling of Pure Substances Electrode Materials}
We start with thermodynamic modeling of pure-substances electrode materials, e.g. Li metal, graphite, etc, with the classical CALPHAD-type framework, but statistical thermodynamics derived expressions instead of commonly-used polynomials\cite{Dinsdale1991} thanks to the differentiability of the thermodynamic modeling framework which allows differentiation through integration via the quadrature. Assuming no other significant terms caused by external fields, the molar Gibbs free energy of a pure substance can be written as\cite{Liu_Wang_2016}
\begin{equation}
    g = h_{static} + g_{vib}(T) + g_{elec}(T)
\end{equation}
where $h_{static}$ is the molar total potential energy of the system at 0K and is a constant. $g_{vib}$ is the vibrational contribution to total Gibbs free energy due to phonon modes, which is temperature dependent and can be decomposed as $g_{vib} = h_{vib}- Ts_{vib}$. $g_{elec} = h_{elec}- Ts_{elec}$ is the electronic contribution to the Gibbs free energy, which is important for metals at high temperatures,\cite{Liu_Wang_2016} and is not discussed further in this work. See supporting information for a derivation of the expression of $h_{elec}$ and $s_{elec}$. 

We first derive the expression of $g_{vib}$, while $g_{static}$ will be discussed in the next section when discussing mixing enthalpy. It should be emphasized that for any molar Gibbs free energy expression $g = h-Ts$, $\frac{\partial h}{\partial T} = T\frac{\partial s}{\partial T}$ constraint must be satisfied.  It is obvious that  $\frac{\partial h_{vib}}{\partial T} = T\frac{\partial s_{vib}}{\partial T}$ must be satisfied. Here, the expression of $h_{vib}$ is given as \cite{Pande2018}
\begin{equation}
    h_{vib} = \int_0^\infty g(\omega) d\omega \left[ \frac{1}{2}\hbar\omega + \frac{\hbar\omega}{e^{\frac{\hbar\omega}{k_BT}}-1} \right]
    \label{eqn:h_vib}
\end{equation}
and $s_{vib}$ as 
\begin{equation}
    s_{vib} = k_B \int_0^\infty g(\omega) d\omega \left[ \frac{\frac{\hbar\omega}{k_BT}}{e^{\frac{\hbar\omega}{k_BT}}-1} - \log(1-e^{-\frac{\hbar\omega}{k_BT}})\right]
    \label{eqn:s_vib}
\end{equation}
where $g(\omega)$ is the phonon density-of-states (PDOS) as a function of phonon frequency $\omega$. The two expressions satisfy $\frac{\partial h_{vib}}{\partial T} = T\frac{\partial s_{vib}}{\partial T}$. To estimate the unknown PDOS $g(\omega)$, we applied a modified Debye model  
\begin{equation}
    g(\omega) = \omega^2 \sum_{i=0}^n g_iP_i(2\frac{\omega}{\omega_{max}}-1)
    \label{eqn:PDOS_x_independent}
\end{equation}
where $g_i$ are fitting parameters, $P_i$ is $i$-th order Legendre polynomial, and the PDOS $g(\omega)$ subjects to the constraint that $\int_0^{\omega_{max}} g(\omega) d\omega = 3N_A$. By using the above modified Debye model, we ensured that PDOS $g(\omega) \propto \omega^2$ at low frequencies for solids. To determine the fitting parameters $g_i$, we utilize the fact that the constant pressure specific heat capacity $c_p = \frac{\partial h_{vib}}{\partial T}$ assuming $h_{mix}$ has weak temperature dependency and is neglected here, therefore the values of  $g_i$ can be determined by fitting $\frac{\partial h_{vib}}{\partial T}$ to experimental measured $c_p$ data. Here, leveraging the power of differentiable physics,\cite{diffphys_position_piece, Kochkov2021_MichaelBrenner} Equation \ref{eqn:h_vib}-\ref{eqn:PDOS_x_independent} are implemented in differentiable programming framework PyTorch, and the parameter of $g_i$ are fitted to $c_p$ by utilizing the automatic differentiation (autograd) of $h_{vib}$ with respect to temperature $T$. Gauss-Legendre quadrature within $[0, \omega_{max}]$ is applied to evaluate Equation \ref{eqn:h_vib}, where $\omega_{max}$ is the maximum frequency of all phonon modes and corresponds to the Debye temperature of the material.  Gradient based optimization of  $g_i$ is performed via minimizing the squared-error loss function between measured and modeled specific heat.  

\begin{figure}[!ht]
    \centering
    \includegraphics[width=8.7cm]{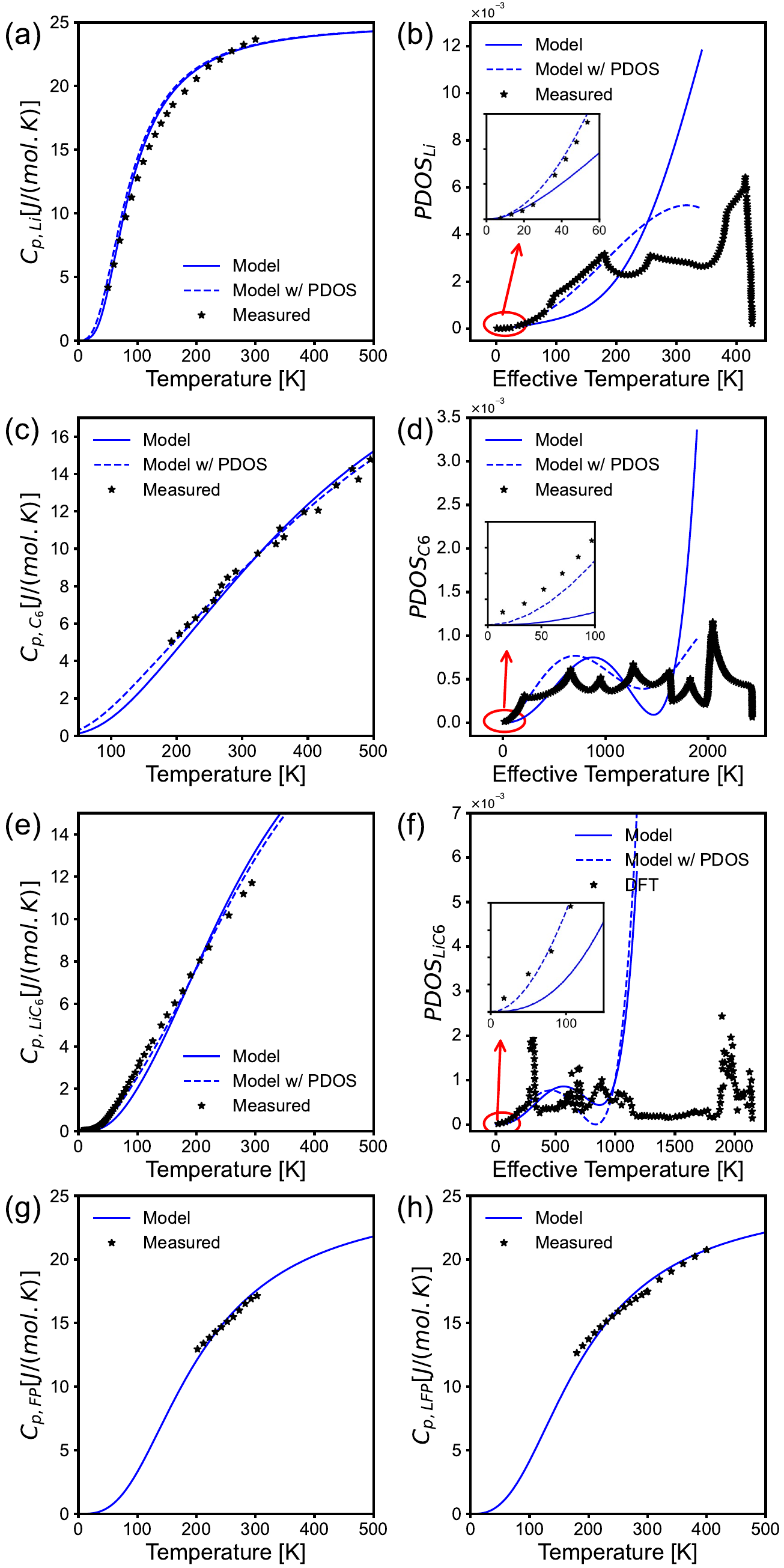}
    \caption{Modeled (blue lines, solid lines represent results fitted purely with specific heat ($c_p$) measurements, dashed lines represent results fitted with both $c_p$ and measured PDOS data) and measured (black dots, experimental measurements or DFT calculation) 
    (a) constant pressure specific heat ($c_p$) of Li metal, (b) fitted PDOS of Li metal (inserted figure shows the low frequency region), effective temperature is calculated by the energy of phonon divided by $k_BT$, (c) constant pressure specific heat ($c_p$) of graphite (\ce{C6}), (d) fitted PDOS of graphite (inserted figure shows the low frequency region), (e) $c_p$ of lithiated graphite (\ce{LiC6}), (f) fitted PDOS of LiC$_6$ (inserted figure shows the low frequency region), note that the black dots here are density functional theory (DFT) calculated by Gianni et al,\cite{LiC6_PDOS} (g) $c_p$ of FP (\ce{FePO4}), and (h) $c_p$ of LFP (\ce{LiFePO4}).  Measured specific heat of graphite is digitized from Butland et al,\cite{graphite_cp}, Li metal from Simon et al,\cite{Li_Cp} \ce{LiC6} from Ayache et al,\cite{Ayache1980}, \ce{FP} from Shi et al,\cite{FP_Cp} \ce{LFP} from Loos et al,\cite{LFP_Cp} PDOS of Li digitized from Beg et al\cite{Beg1976}, PDOS of graphite digitized from Young and Koppel,\cite{graphite_cp} DFT-calculated PDOS digitized from Gianni et al.\cite{LiC6_PDOS} PDOS curves are normalized such that the areas under PDOS curves equal to 1. }
    \label{fig:pure_substances}
\end{figure}

As demonstration, we conducted thermodynamic modeling of \ce{Li} metal anode, graphite (\ce{C6}), lithiated graphite \ce{LiC6}, iron phosphate \ce{FePO4}, lithium iron phosphate \ce{LiFePO4}, and compare with experimental measurements. As shown with the solid blue lines in Figure \ref{fig:pure_substances}(a),(c), (e), (g)-(h), the fitted $h_{vib}$ for the five pure substances lead to accurate reproduction of constant pressure specific heat, with the average absolute error to be 0.71 $J/(mol\cdot K)$ for Li, 1.24 $J/(mol\cdot K)$ for graphite, 0.33 $J/(mol\cdot K)$ for \ce{LiC6}, 0.34 $J/(mol\cdot K)$ for FP, and 0.43 $J/(mol\cdot K)$ for LFP. Note that the value of $g_i$ in PDOS expression of Li metal is fitted using data points beyond 220K, ensuring all measured data of $c_p$ of Li is in bcc phase rather than fcc phase's.\cite{Li_bcc_fcc} Since the value of $g_i$ in the PDOS expression are fitted with $c_p$ data, we also plotted the modeled PDOS of the Li in Figure \ref{fig:pure_substances} (b) with blue solid line, and compared with measured PDOS of Li\cite{Beg1976} with neutron scattering. It is noticed that although the fitted $c_p$ demonstrate good accuracy, the PDOS of these pure substances are not captured well, except that the modified Debye model is guaranteed to satisfy $g(\omega) \propto \omega^2$ at low frequencies since the leading term in the expression is proportional to $\omega^2$, as shown in the insert figure of Figure \ref{fig:pure_substances} (b). Further improvement of PDOS modeling can be done by introducing additional measurements that give more information than $c_p$, e.g. one way is to include experimental measurements of PDOS\cite{IAEA_Li_PDOS, graphite_PDOS} or density functional theory calculated PDOS\cite{LiC6_PDOS} as constraint and fit the PDOS expression (i.e. Equation \ref{eqn:PDOS_x_independent}) together with $c_p$ measurements, such methods are applied to Li, graphite and LiC$_6$ and are demonstrated with blue dashed lines in Figure \ref{fig:pure_substances} (a)-(f). Compared with the PDOS fitted only with $c_p$ measurements, the PDOS fitted with both $c_p$ and PDOS measurements in general reproduce the measured PDOS more accurately as shown in Figure \ref{fig:pure_substances} (b), (d) and (f), and especially in the low-frequency region as shown in the inserted figures. Although being possibly achievable with additional measurements, more accurate determination of PDOS is beyond the scope of this paper, since the primary goal of this work is to model the thermodynamic variables (e.g. $h_{vib}$, etc) and thermodynamic properties (e.g. $c_p$, etc) that are useful for analyzing the thermodynamics of electrochemical energy storage systems.

\section*{Thermodynamic Modeling of Mixed Substances}
Using the modeled Gibbs free energy of pure substances, the expression of Gibbs free energy of mixed substances can be  discussed.  For electrode material (HM) of Li-ion batteries, the lithiation process of HM can be viewed as mixing pure substance of Li and pure substance HM, resulting in a mixed substance \ce{Li_xHM}. Therefore, we can write the total molar Gibbs free energy of mixed substance, i.e. lithiated host material, \ce{Li_xHM} as 

\begin{equation}
    g_{\ce{Li_x HM}} = xg_{\ce{Li}} + g_{\ce{HM}} + g_{excess} 
    \label{eqn:g_excess}
\end{equation}
where the excess Gibbs free energy $g_{excess}$ can be decomposed as $g_{excess} = h_{excess} - Ts_{excess}$. According to different contributing sources, $h_{excess}$ can be decomposed into
\begin{equation}
    h_{excess} = h_{excess, mix}(x) + h_{excess, vib}(x,T) + h_{excess, elec}(x,T)
    \label{eqn:h_excess}
\end{equation}
where the excess mixing enthalpy 
\begin{equation}
    h_{excess, mix}(x) = h_{static, \ce {Li_x HM}} - h_{static, \ce{HM}} -  x\cdot h_{static, \ce{Li}}
\end{equation}
arises arises from Li-ion interactions in lithiated host material \ce{Li_x HM} that depends on composition $x$, and can be expanded as a series of polynomials. Previously, Yao and Viswanathan\cite{yao2024open} expanded the excess mixing Gibbs free energy using Redlich-Kister polynomials.\cite{redlich1948algebraic} The history of expanding excess free energy with polynomials can be traced back to Margules in 1895.\cite{gokcen1996gibbs} Later in 1936, Guggenheim\cite{guggenheim1937theoretical} proposed using polynomial series of the form $x(1-x)\sum_{i=0}^n A_i (1-2x)^i$ to expand excess free energy of a binary system. Later in 1948, Redlish and Kister\cite{redlich1948algebraic} proposed to use the expression proposed by Guggenheim in a multicomponent system by adding the expressions for all binary pairs,\cite{scatchard1949equilibrium} whose method is referred to as Redlish-Kister expansion. In 1974, Bale and Pelton\cite{bale1974mathematical} showed that using Legendre polynomials to expand the excess Gibbs free energy leads to a decoupling of the coefficients, which makes it more amenable to physical interpretation.  This is because of the orthonormal property of Legendre polynomials: $\int_0^1 P_n(1-2x) P_m(1-2x) d(1-2x) = \delta_{nm}$. Here, in the same spirit of Bale and Pelton's works,\cite{bale1974mathematical, pelton1986legendre} we expand $h_{excess, mix}(x)$ as  $h_{excess, mix}(x) = G_0 + x(1-x) \sum_{i=0}^n \Omega_i P_i(1-2x)$,
where $P_i$ is the i-th order Legendre polynomials, $\Omega_i$ and $G_0$ are the coefficients that can be determined by fitting to experimentally measured OCV data. 

The excess vibrational enthalpy
\begin{equation}
    h_{excess, vib}(x,T) = h_{vib, \ce{Li_x HM}} - h_{vib, \ce{HM}} - xh_{vib, \ce{Li}}
    \label{eqn:h_excess_vib}
\end{equation}
is the change of vibration (phonon) enthalpy, and is a function of both composition $x$ and temperature $T$. $h_{excess, elec}(x,T)$ is the change of electronic enthalpy after mixing \ce{Li} into \ce{HM} to form \ce{Li_x HM}. As $HM$ and $Li$ are pure substances, the expression of $h_{vib, \ce{HM}}$ and $h_{vib, \ce{Li}}$ can be fitted with the measured constant pressure specific heat data, assuming $h_{mix}$ has weak temperature dependency and are neglected here. The vibrational enthalpy of lithiated host material \ce{Li_x HM} can also be written using Equation \ref{eqn:h_vib}, however the PDOS is Li-filling-fraction ($x$) dependent. In line with the previous analysis, we expand the PDOS of \ce{Li_x HM} as 
\begin{equation}
    g(\omega, x) = \omega^2 \sum_{i=0}^n g_i(x)\cdot P_i(2\frac{\omega}{\omega_{max}}-1)
    \label{eqn:PDOS_x_dependent}
\end{equation}
where
\begin{equation}
    g_i(x) = \sum_{j=0}^m g_{ij} \cdot P_i(2x-1)
    \label{eqn:PDOS_x_dependent_gij}
\end{equation}
$P_i(2x-1)$ is Legendre polynomials of i-th order, $g_{ij}$ are fitting parameters. As discussed before, we assume $h_{excess, elec}(x,T)$ is negligible.

Similar to how excess enthalpy is decomposed, we can write the excess entropy as 
\begin{equation}
    s_{excess} = s_{excess, mix}(x) + s_{excess, vib}(x,T) + s_{excess, elec}(x,T)
\end{equation}
where $s_{excess, mix}(x)$ is the configurational entropy that arises from mixing \ce{Li} into \ce{HM}, and can be approximated by the ideal mixing entropy $-k_B \left[ x\log x + (1-x) \log(1-x)\right]$.\cite{yao2024open} In this work, in addition to the ideal mixing entropy, we account for the fact that not all Li-filling sites are available during fitting,\cite{Pande2018} and the expression of $s_{excess, mix}(x)$ is written as 
\begin{equation}
    s_{excess, mix}(x) = -k_B \left[ x\log x + (1-x) \log(1-x)\right] C
    \label{eqn:s_mix}
\end{equation}
where the non-ideal correction term $C =  (1+\omega_0) + \sum_{i=1}^n \omega_i P_i(1-2x)$,  $P_i(x)$ is Legendre polynomials of i-th order. We used $1-2x$ to ensure $P_i(1-2x)$ is orthonormal basis in $(1-2x) \in [0,1]$.  $\omega_i$ are fitting parameters, which value can be fitted from the measured entropic heat data of \ce{Li_x HM} after $s_{vib}$ are determined. Similar to how PDOS parameters are determined in the previous section, automatic differentiation can be applied to Equation \ref{eqn:s_vib} and \ref{eqn:s_mix} to calculate $\frac{ds}{dx}$, and then gradient based optimization of $\omega_i$ is performed via minimizing the squared-error loss function between measured and modeled entropic heat.

Similar to how excess vibrational enthalpy is calculated, we write excess vibrational entropy as 
\begin{equation}
    s_{excess, vib}(x,T)= s_{vib, \ce{Li_x HM}} - s_{vib, \ce{HM}} - xs_{vib, \ce{Li}}
    \label{eqn:s_excess_vib}
\end{equation}
where $s_{vib, \ce{Li_x HM}}$, $ s_{vib, \ce{HM}}$, and $s_{vib, \ce{Li}}$ are expressed as Equation \ref{eqn:s_vib}, and the PDOS of \ce{Li_x HM} is expressed as Equation \ref{eqn:PDOS_x_dependent}. In this work, we neglect the effect of $s_{excess, elec}(x,T)$ and $h_{excess, elec}(x,T)$ due to the temperature range discussed in this work. 

Having the excess molar Gibbs free energy $g_{excess}$ modeled as above, it is natural to derive the expression of OCV of the host material \ce{HM}. Consider the reaction of Li-ion intercalating into host material \ce{HM} and occupying a site in \ce{HM}, i.e. before reaction there is a Li-ion, an electron and an unoccupied site in \ce{HM}, and after reaction there is an occupied site in \ce{HM}:
\begin{equation}
    \ce{Li+ + e- + HM = Li-HM}
\end{equation}
Enforcing equilibrium indicates the chemical potential of two phases must be at equilibrium, i.e. the summation of chemical potential at left hand side equals to that at right hand side.
\begin{equation}
    \mu_{\ce{Li+}} + \mu_{\ce{e-},\ce{HM}} + \mu_{\ce{HM}} = \mu_{\ce{Li-HM}}
\end{equation}
Referencing to the Li metal anode,
\begin{equation}
    \mu_{\ce{Li+}} + \mu_{\ce{e-},ref}= \mu_{\ce{Li}}
\end{equation}
we have
\begin{equation}
    \mu_{\ce{Li}} + (\mu_{\ce{e-},\ce{HM}}-\mu_{\ce{e-}, ref}) + \mu_{\ce{HM}} = \mu_{\ce{Li-HM}}
\end{equation}
plugging-in Nernst equation,
\begin{equation}
    \mu_{\ce{Li}} -e(U_{HM}-U_{ref})+ \mu_{\ce{HM}} = \mu_{\ce{Li-HM}}
\end{equation}  
\begin{equation}
    \mu_{\ce{Li}} -e\times OCV + \mu_{\ce{HM}} = \mu_{\ce{Li-HM}}
\end{equation}     
Therefore 
\begin{equation}
     -e\times OCV  = (\mu_{\ce{Li-HM}} - \mu_{\ce{HM}}) - \mu_{\ce{Li}}
     \label{eqn:OCV}
\end{equation} 
i.e. OCV is proportional to the chemical potential difference between occupied and unoccupied site in \ce{HM}, subtracted by bulk Li chemical potential. Or equivalently, OCV is proportional to the "excess" chemical potential of mixing \ce{Li} and \ce{HM}, i.e.
\begin{equation}
     -e\times OCV  = \mu_{\ce{Li-HM}} - \mu_{\ce{HM}} - \mu_{\ce{Li}} = \frac{\partial g_{excess}}{\partial x} 
     \label{eqn:OCV_gexcess}
\end{equation} 
where $g_{excess}$ is defined in Equation \ref{eqn:g_excess} and can be calculated according to the above discussions. It should be noted that as Yao and Viswanathan pointed out,\cite{yao2024open}, the OCV must be a monotonic function of $x$ as ensured by the second law of thermodynamics. Non-monotonic region of calculated OCV indicates phase coexistence, and OCV within this region can be re-calculated as the coexistence chemical potential divided by the negative Faraday constant. In this work, all the above expressions of Gibbs free energy are implemented on top of the differentiable thermodynamic modeling code in Yao and Viswanathan,\cite{yao2024open} which allows calculation of Gibbs free energy of substances, and OCV and specific heat via automatic differentiation. Finally, a phase diagram can be plotted given the phase coexistence regions have been calculated within the differentiable thermodynamics framework. 

\begin{figure}
    \centering
    \includegraphics[width=8.7cm]{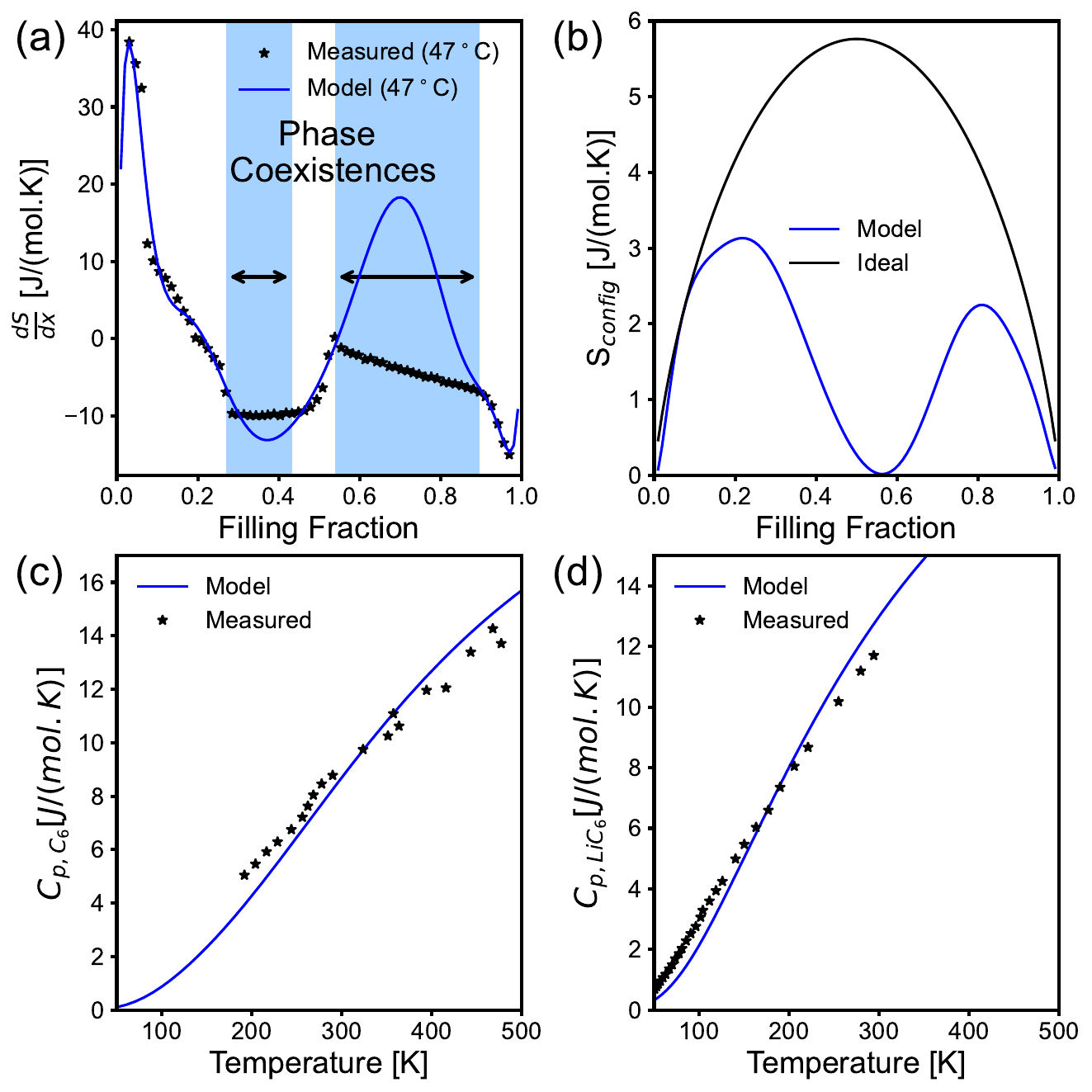}
    \caption{(a) Fitted $\frac{ds}{dx}$ from the fitted graphite OCV model (blue line) at 47$^\circ$C compared with experimental measured data (black dots)\cite{Mercer_graphite2021}, the phase coexistence regions are determined by the differentiable thermodynamics framework (in Figure \ref{fig:OCV} and are highlighted as blue, (b) fitted molar configurational entropy $s_{total}$ (blue line) and ideal entropy (black line), $c_p$ of (c) graphite and (d) \ce{Li_1 C6} modeled by the fitted PDOS of \ce{Li_x C6} at $x=0$ and $x=1$ given by Equation \ref{eqn:PDOS_x_dependent} and \ref{eqn:PDOS_x_dependent_gij} (blue lines) compared with measured $c_p$ (black dots). }
    \label{fig:thermo_of_LixC6}
\end{figure}

As demonstration, we performed thermodynamic modeling of \ce{Li_x C6}, one of the most popular anode materials for Li-ion batteries. We first examine the fitted $s_{vib}$ and $s_{excess,mix}$ of \ce{Li_x C6} as shown in Figure \ref{fig:thermo_of_LixC6}. Figure \ref{fig:thermo_of_LixC6}(a) shows the fitted $\frac{ds}{dx}$ compared with experimental measurement by Mercer et al,\cite{Mercer_graphite2021} indicating the fitted $g_{i,j}$ and $\omega_i$ values reproduce the experimental results well. Within the phase coexistence regions as determined by the differentiable thermodynamics framework, the value of $\frac{ds}{dx}$ is determined by the two end points of phase coexistence regions\cite{baek2022thermodynamic} and thus the fitted model does not need to be close to the measured data points. Figure \ref{fig:thermo_of_LixC6}(b) shows the fitted configurational molar entropy $s_{config}$ of graphite comparing with the ideal mixing entropy $s_{ideal} = -k_B \left[ x\log x + (1-x) \log(1-x)\right]$.  It is found that the fitted configurational entropy of graphite is always smaller than the ideal mixing entropy. At around filling fraction $x=0.5$, the local minima of configurational entropy indicates local ordering of Li-ions within graphite, such observation is consistent with previous reports that a stage-2 lithiated graphite forms, i.e. a layer with all empty sites occupied by Li followed by another layer with no Li occupation.\cite{Mercer_graphite2021, Pande2018} Since the value of $g_{i,j}$ is the same for vibrational entropy $s_{vib}$ and vibrational enthalpy $h_{vib}$ of \ce{Li_x C6}, we examine the quality of fitted $g_{i,j}$ again using the calculated constant pressure specific heat $c_p$ and comparing with experimental measurements, as shown in Figure \ref{fig:thermo_of_LixC6} (c) for graphite (i.e. when $x=0$ for \ce{Li_x C6}), and (d) for \ce{Li C6} (i.e. when $x=1$ for \ce{Li_x C6}). It is found that the $c_p$ modeled by the fitted PDOS of \ce{Li_x C6} reaches average absolute error of 1.31 $J/(mol \cdot K)$ for graphite and 1.22 $J/(mol \cdot K)$ for \ce{Li C6}, thus ensuring good quality of the fitted $g_{ij}$.

\begin{figure}[h!]
    \centering
    \includegraphics[width=8.7cm]{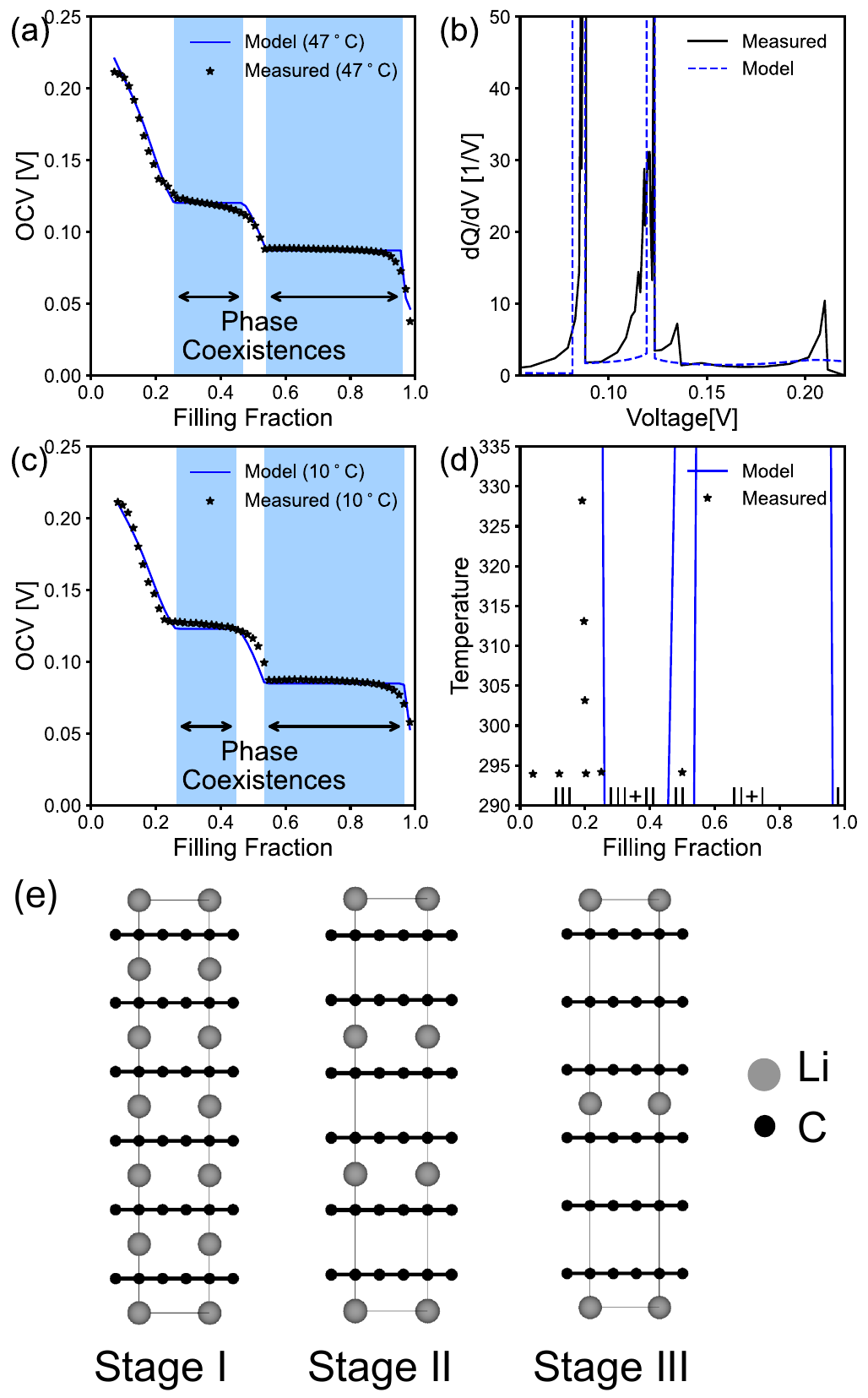}
    \caption{(a) Fitted graphite OCV at 47$^\circ$C (blue line) compared with experimental measured data (black dots)\cite{Mercer_graphite2021}, the phase coexistence regions are determined by the differentiable thermodynamics framework and are highlighted as blue, (b) dQ-dV curve calculated from fitted graphite OCV model at 47$^\circ$C (blue line) compared with that calculated from experimental measured data (black dotted line)\cite{Mercer_graphite2021}, (c) graphite OCV extrapolated to 10$^\circ$C (blue line) compared with experimental measured data (black dots)\cite{Mercer_graphite2021}, (d) calculated phase diagram of \ce{Li_x C6} (blue lines) compared with the measured phase boundaries of (black dots) by Dahn.\cite{Dahn1991}, and (e) schematic diagram (side view) of Stage I, II and III lithiated graphite.}
    \label{fig:OCV}
\end{figure}

Next, we examine the fitted OCV of \ce{Li_x C6}. Since we expressed the OCV of \ce{Li_x C6} as the partial derivative of $g_{excess}$ with respect to $x$, while $g_{excess}$ is written as a function of both $x$ and temperature $T$, therefore it is convenient for us to model OCV of \ce{Li_x C6} as function of both $x$ and $T$. Figure \ref{fig:OCV}(a) shows the OCV calculated from thermodynamic modeling compared with experimental measurements at 47$^\circ$C. The MAE of OCV model with respect to experimental measurements is 2.90 mV. Differential voltage analysis of the fitted OCV model shown in Figure \ref{fig:OCV} (b) showed that the fitted graphite OCV model is able to capture the peaks at around around 0.08V and 0.12V, indicating the ability to capture the phase change. Indeed, since the differentiable thermodynamic modeling guarantees that any concave region in Gibbs free energy curve is automatically detected and modeled as phase coexisting region, whose coexisting chemical potential is determined by solving common tangent condition, such ability to capture phase change is guaranteed by construction.  Since the fitted graphite OCV model is a function of temperature, we also evaluated its accuracy at other temperatures. At 25$^\circ$C, MAE is 3.00 mV, while at 57$^\circ$C, MAE is 3.64 mV. It should be emphasized that the fitted graphite OCV model is extrapolated to 10$^\circ$C as shown in Figure \ref{fig:OCV}(c), a temperature that is not included in the fitting data, and the graphite OCV model reaches MAE of 3.81 mV. As we have already constructed molar Gibbs free energy  of \ce{Li_x HM} $g_{Li_xHM}$ from the above thermodynamic analysis, it is natural to produce the phase diagram of \ce{Li_x HM} using the constructed $g_{Li_xHM}$. The constructed phase diagram of \ce{Li_x HM}  is shown in Figure \ref{fig:OCV}(d), which clearly shows the three stages of graphite (marked as III, II and I) and their coexistence (III+II, II+I) during Li intercalation and is consistent with previous analysis.\cite{Dahn1991, Pande2018} 

With the constructed $g_{Li_xHM}$ and its mixing contributions $h_{mix}$ and $s_{mix}$, as well as vibration contributions $h_{vib}$ and $s_{vib}$, we looked into the slope of phase boundary as a function of temperature in Figure \ref{fig:OCV}(d). For the phase boundary between III and III+II, its negative slope is mainly dominated by the effect of configurational entropy stabilization. As temperature increases, the configurational entropy term $-Ts_{mix}$ in the Gibbs free energy expression of $Li_xC_6$ increases. The lithium stoichiometry $x$ of the stage III compound in the III+II coexistence region moves towards the direction that has a larger $s_{mix}$ which decreases the Gibbs free energy of the III+II region, and given the $s_{mix}$ as a function of $x$ was retrieved from entropic heat measurement as shown in Figure \ref{fig:thermo_of_LixC6}(b), the peak value of $s_{mix}$ is at x=0.21 which is smaller than the $x=0.24$ at III and III+II boundary, therefore the lithium stoichiometry $x$ of the stage III compound in the III+II coexistence region moves towards negative x direction as temperature increases, resulting in the negative slope of the phase boundary between III and III+II. The effects of vibrational contribution as well as mixing enthalpy contribution are relatively smaller at the boundary between III and III+II, as discussed in the supporting information. For all other phase boundaries, the slopes are determined by the competition between configurational entropy $s_{mix}$ which moves towards the direction of increasing itself, and mixing enthalpy $h_{mix}$ which tends to minimize itself. Notably, such explanations on the behavior of phase boundary can be only obtained with the electrochemical measurements of entropic heat which allows accurate determination of configurational entropy instead of using ideal configurational entropy used typically in battery modeling.\cite{yao2024open, baek2022thermodynamic} 


It is worth mentioning that OCV and entropic heat measurement at finite C rate and lack of reported measurement uncertainty pose challenges to uncertainty quantification of the constructed phase diagram.\cite{OTIS2022111590}  Bayesian differentiable thermodynamics can be done in the future for uncertainty quantification of the thermodynamic modeling of electrode materials.


\section*{The Inverse Problem of Thermodynamic modeling: Learning OCV from Phase Diagrams}
\begin{figure}[h!]
    \centering
    \includegraphics[width=8.7cm]{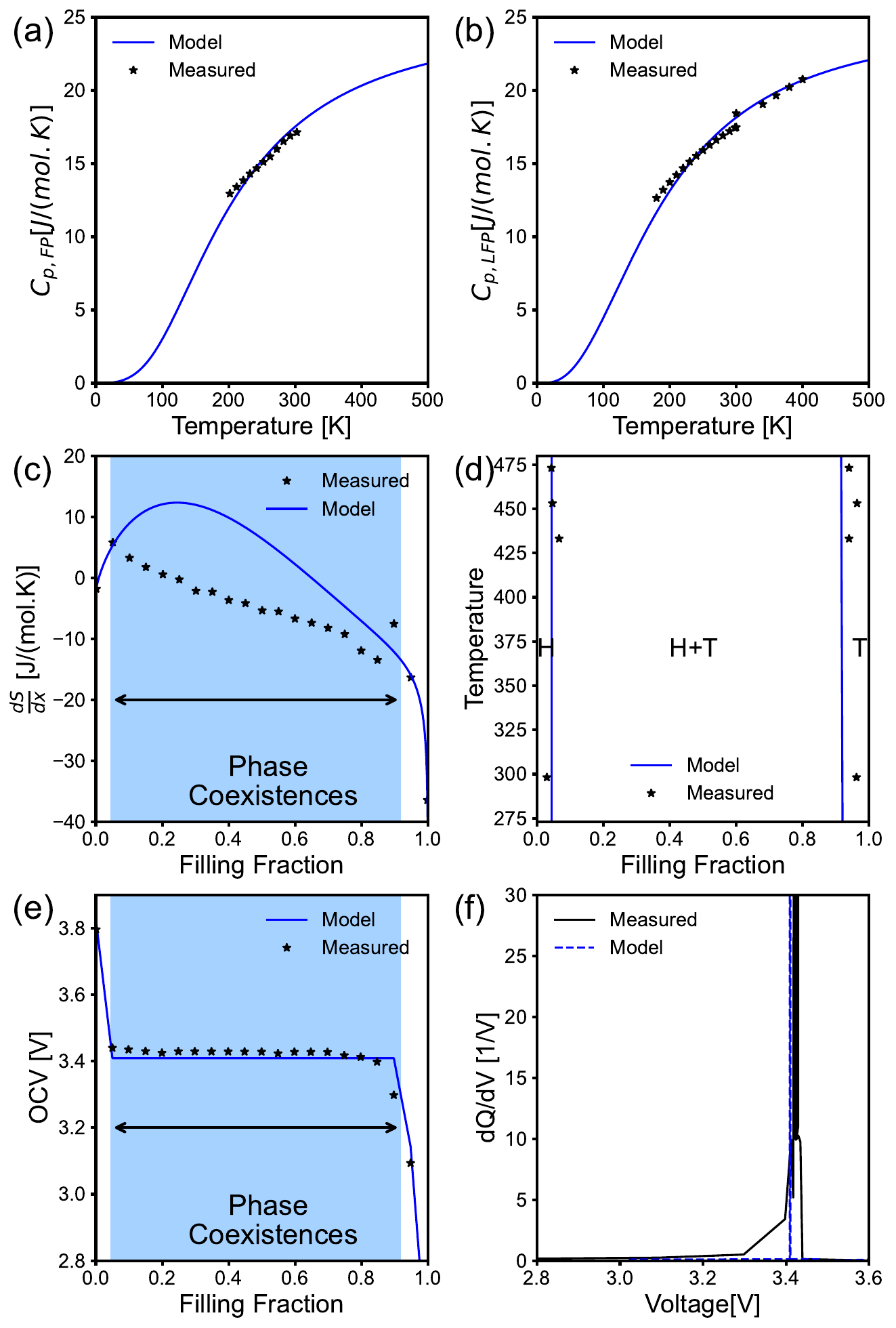}
    \caption{$c_p$ of (a) FP and (d) LFP modeled by the fitted PDOS of \ce{Li_x FePO4} at $x=0$ and $x=1$ given by Equation \ref{eqn:PDOS_x_dependent} and \ref{eqn:PDOS_x_dependent_gij} (blue lines) compared with measured $c_p$ (black dots). (c) Fitted $\frac{ds}{dx}$ from the fitted LFP OCV model (blue line)  compared with experimental measured data (black dots)\cite{Dodd_Thesis}, the phase coexistence regions are determined by the differentiable thermodynamics framework and are highlighted as blue, (d) calculated phase diagram of LFP (blue lines) compared with experimental measured phase boundaries under 200$^\circ$C from Dodd et al(black dots).\cite{Dodd2006_PD} (e) Fitted LFP OCV (blue line) compared with experimental measured data (black dots)\cite{Dodd_Thesis}, the phase coexistence regions are determined by the differentiable thermodynamics framework and are highlighted as blue, (f) dQ-dV curve calculated from fitted LFP OCV model (blue line) compared with that calculated from experimental measured data (black dotted line)\cite{Dodd_Thesis}.}
    \label{fig:LFP}
\end{figure}

The differentiability of thermodynamic modeling framework enables the inverse problem, i.e. learning Gibbs free energy from phase equilibrium data (e.g. phase boundaries in phase diagram). In principle, phase boundaries are implicit functions of the thermodynamic parameters (i.e. $g_{ij}, \omega_i, \Omega_i, G_0$, etc.) in the Gibbs free energy expressions. However, optimization of the parameters is hard since no explicit expressions between these parameters and phase boundaries can be written out. In this work, since the phase boundaries can be calculated in the differentiable thermodynamics, the gradient of phase boundaries $x$ with respect to thermodynamic parameters $\theta$, i.e.$\frac{dx}{d\theta}$ can be calculated with automatic differentiation, thus enabling gradient-based optimization of $\theta$ according to phase boundaries $x$. Previously, Guan has demonstrated learning Gibbs free energy expressions of a binary alloy from purely phase equilibrium data with differentiable thermodynamics.\cite{Guan2022} Here, in addition to phase equilibrium data of LFP reported by Dodd,\cite{Dodd_Thesis} we construct Gibbs free energy expression of LFP from thermochemistry data (e.g. measured $c_p$ of Li,\cite{Li_Cp} FP,\cite{FP_Cp} LFP\cite{LFP_Cp}) and electrochemistry data (OCV and $\Delta S$).\cite{Dodd_Thesis} The results are summarized in Figure \ref{fig:LFP}. Same as previous sections, we first evaluate the fitted $c_p$ of pure substances, i.e. FP and LFP, in Figure \ref{fig:LFP} (a) and (b) respectively. The average absolute error of fitted $c_p$ with respect to measured ones are 0.36 J/(mol$\cdot$K) for FP, and 0.38 J/(mol$\cdot$K) for LFP. The fitted $\Delta S$, i.e. $\frac{ds}{dx}$ or equivalently $\frac{dOCV}{dT}$, is examined in Figure \ref{fig:LFP}(c). The measured phase boundaries of LFP at different temperatures\cite{Dodd_Thesis} shown in Figure \ref{fig:LFP}(d) is taken as input data in addition to $c_p$, $\frac{ds}{dx}$ and OCV data to model the Gibbs free energy expression, and the phase boundary calculated from the modeled Gibbs free energy is shown in the same figure for comparison.  
Analysis on the slope of phase boundaries between H and H+T as well as T and H+T can be drawn similar to that for the graphite case as mentioned in the previous section, i.e. competition between $s_{mix}$ and $h_{mix}$ decides the slope of the phase boundaries.
In general, the modeled Gibbs free energy accurately captures the phase boundaries. As a result, the OCV is accurately modeled as shown in Figure \ref{fig:LFP}(e), with MAE of 0.0276 V. The dQ-dV curve calculated from fitted LFP OCV model is compared with that calculated from experimental measured OCV of LFP and is shown in Figure \ref{fig:LFP}(f). As a consequence of correctly capturing the phase coexistence, the modeled dQ-dV curve correctly captures the peak around 3.4V.  

\section*{Summary}
In summary, we propose a differentiable thermodynamic modeling framework for electrode materials of electrochemical systems. Within the proposed framework, electrochemical data, such as OCV and entropic heat which contain rich thermodynamic information, are unified with thermochemical data, such as phase equilibrium and specific heat. We also demonstrate integration with the phonon density of states $g(\omega)$ through the modified Debye model, which captures the $g(\omega) \propto \omega^2$ dependency in the low-frequency region and enables thermodynamically consistent models from microscopic to macroscopic information.  To realize this novel capability within thermodynamic modeling, we integrate via Gauss-Legendre quadrature, and leverage differentiable programming to perform gradient-based learning of model parameters from all kinds of thermochemical and electrochemical data.  With the broadened range of data, we provide thermodynamically consistent models for phonon density of states, specific heat, configurational entropy, OCV, and phase diagrams for graphite and LFP, two popular electrode materials. In particular, entropic heat data recovers configurational entropy, thus unlocking insights into the slope of phase boundaries of graphite and LFP phase diagrams, which is impossible to achieve without entropic heat data. 
Given the capability of building thermodynamically consistent model that captures various macroscopic (OCV, specific heat) and microscopic (configurational entropy, PDOS) thermodynamic properties of electrode materials, we believe this work can unlock a deeper and seamless bridge between classical thermodynamic modeling and battery modeling.

\acknow{This work was supported in part by the Defense Advanced Research Projects Agency (DARPA) under contract no. HR00112220032. The results contained herein are those of the authors and should not be interpreted as necessarily representing the official policies or endorsements, either expressed or implied, of DARPA, or the U.S. Government. }

\showacknow{} 


\bibliography{ref}

\end{document}